\documentclass[12pt]{iopart}

\usepackage{amssymb,graphicx}

\newcommand{\fig}[4][h]{\begin{figure}[#1]\begin{center}\includegraphics[width=#2]{figures/#3.pdf}\vspace{-0.25 cm}\caption{#4}\label{fig:#3}\end{center}\end{figure}}

\newcommand{\OmegaR}{\ensuremath{\Omega_{\rm{R}}}}

\newcommand{\totalD}[2]{\frac{\rm{d} #1}{\rm{d} #2}}
\newcommand{\eqref}[1]{(\ref{#1})}

\def\ensuretext{\textrm}

\newcommand{\ket}[1]{\ensuremath{\left|{#1}\right\rangle}}

\newcommand{\Rb}[1]{\ensuremath{^{#1}}\ensuretext{Rb}}
\newcommand{\hf}[2]{\ket{#1, \; #2}}
\newcommand{\units}[1]{\ensuretext{\thinspace #1}}

\begin{document}

\title[Optically trapped atom interferometry using the clock transition of large \Rb{87} BECs]{Optically trapped atom interferometry using the clock transition of large \Rb{87} Bose-Einstein condensates}

\author{P.\,A. Altin, G. McDonald, D. D\"oring, J.\,E. Debs, T.H. Barter, N.\,P. Robins and J.\,D. Close}
\address{Department of Quantum Science, ARC Centre of Excellence for Quantum Atom Optics, the Australian National University, ACT 0200, Australia}
\author{S.\,A. Haine}
\address{School of Mathematics and Physics, ARC Centre of Excellence for Quantum-Atom Optics, The University of Queensland, Queensland 4072, Australia}
\author{T.\,M. Hanna}
\address{Joint Quantum Institute, National Institute of Standards and Technology and University of Maryland,\\ 100 Bureau Drive, Stop 8423, Gaithersburg, Maryland 20899-8423, USA}
\author{R.\,P. Anderson}
\address{School of Physics, Monash University, Victoria 3800, Australia}

\ead{paul.altin@anu.edu.au}
\date{\today}

\begin{abstract}
We present a Ramsey-type atom interferometer operating with an optically trapped sample of $10^6$ Bose-condensed \Rb{87} atoms. The optical trap allows us to couple the $\hf{F=1}{m_F=0} \rightarrow \hf{F=2}{m_F=0}$ clock states using a single photon 6.8\,GHz microwave transition, while state selective readout is achieved with absorption imaging. Interference fringes with contrast approaching $100\%$ are observed for short evolution times. We analyse the process of absorption imaging and show that it is possible to observe atom number variance directly, with a signal-to-noise ratio ten times better than the atomic projection noise limit on $10^6$ condensate atoms. We discuss the technical and fundamental noise sources that limit our current system, and outline the improvements that can be made. Our results indicate that, with further experimental refinements, it will be possible to produce and measure the output of a sub-shot-noise limited, large atom number BEC-based interferometer. 
\end{abstract}

\submitto{\NJP}
\maketitle


\section{Introduction}
\label{sect:intro}

Atom interferometers have evolved significantly over the last two decades, from systems with matter-based gratings and thermal beams, to systems based on ultra-cold atoms and optical or microwave transitions \cite{cronin09}. Atom interferometers are the basis of our time standard, in the form of laser-cooled Cs fountain clocks with accuracies of a few parts in $10^{16}$ \cite{jefferts02}. The sensitivity of inertial sensors based on atom interferometers is quickly approaching, and in some cases surpassing, state-of-the-art mechanical or optical systems \cite{chu99}. At the same time a number of measurements have proven the competitive nature of atom interferometers in fundamental tests and measurements \cite{lamporesi08}. Perhaps more tantalisingly, it has been proposed that instruments based on atom interferometers can complement the LIGO gravitational wave observatory, allowing access to the $1\units{mHz}$ to $10\units{Hz}$ frequency band \cite{dimopoulos08}.

Laser-cooled atoms have become the source of choice for atom interferometry \cite{cronin09}. The very low centre of mass velocity and narrow energy width of laser-cooled samples allows for a more compact apparatus and better beam-splitting efficiency than hot thermal beam based systems. For optical interferometry, lasers are a very convenient source, offering high flux, low divergence and narrow linewidth. An analogous source for atoms is a Bose-Einstein condensate \cite{anderson95,davis95,bradley95}. In addition to having a dramatically narrower velocity width than a laser cooled source, which facilitates high contrast, large momentum beam splitting, BECs offer the possibility of quadrature squeezing \cite{kitagawa93} as a path to improving the signal-to-noise in an interferometric measurement \cite{mueller08}.  Very recently, an increase in interferometric sensitivity has been demonstrated via quadrature squeezing in small condensed samples \cite{riedel10,gross10}.

The possibilities offered by Bose-condensed sources have driven the development of magnetically trapped clocks and waveguide structures for interferometry with the aim of producing robust portable devices \cite{fortagh07}. These devices typically operate with small atom numbers, N (on the order of $1000-10,000$ atoms), for which projection noise fluctuations are a large fraction of the total atom number ($\sqrt{N}/N \sim 1\,\units{\%}$ to $3\,\units{\%}$) and thus are readily observable. Recently, our group has demonstrated a free space atom interferometer operating at the quantum projection noise limit with $10^4$ atoms \cite{doering10}.

Ideally, one would like to use larger atom clouds for interferometric sensors, since the achievable phase sensitivity increases with number ($\Delta\phi\sim1/\sqrt{N}$). Using condensates containing $10^6$ atoms (readily created in many BEC machines) would immediately give more than an order of magnitude improvement in signal-to-noise over current trapped atom interferometers. The shot noise on such a large cloud is of order $\sqrt{N}/N \sim 0.1\units{\%}$, and detecting at this level puts stringent requirements on all technical aspects of the experimental system. In particular, the interferometer beam-splitters must be extremely stable, and the state-selective measurement of atom number must be very precise. Understanding the technical limitations to these components via measurements and calculations is critical to progress on large condensate atom interferometry.

This paper is divided into several sections: Section \ref{sect:sensitivity} introduces the sensitivity of a Ramsey-type atom interferometer, including a theoretical analysis of quantum projection noise and typical experimental noise sources. Section \ref{sect:interferometer} describes the apparatus, and presents results from our trapped atom interferometer. Section \ref{sect:imaging} discusses the requirements for an absorption imaging system capable of measuring below the shot noise limit on a large atom cloud. We present results on shot noise limited single images as well as analysing the possibility of observing large atom number squeezing with absorption imaging. Section \ref{sect:squeezing} proposes an interferometer with sensitivity enhanced beyond the standard quantum limit that could be implemented with the apparatus described in this paper. We conclude by predicting the achievable signal-to-noise ratio for clocks and inertial sensors based on condensed atom interferometry using the clock states.

\fig{\textwidth}{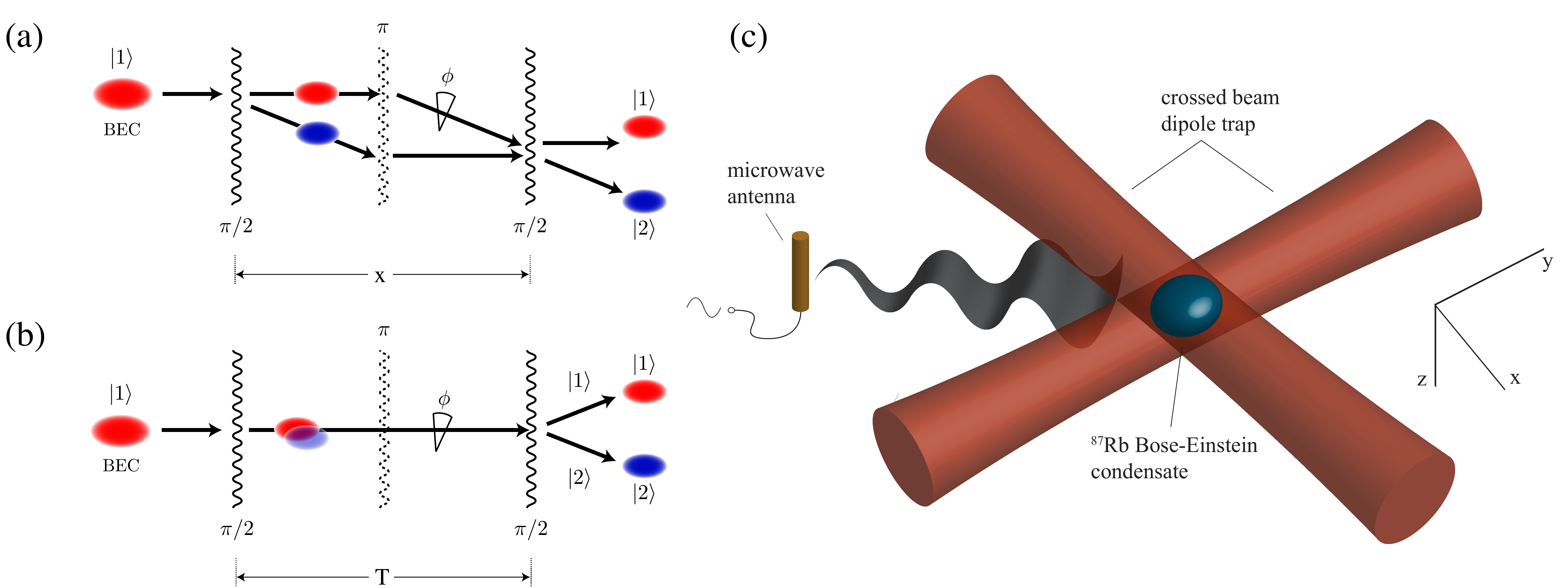}{The Ramsey interferometer scheme. An atomic wavepacket is split into two components, allowed to evolve for a time $T$, and then recombined. The atoms can be coupled to a different internal state, remaining spatially overlapped, or can be coupled to another momentum state, so that the interferometer encloses an area. (a) A spatial Ramsey interferometer is sensitive to inertial effects. (b) The subject of this paper: a temporal Ramsey interferometer is sensitive to state dependent phase shifts. A $\pi$ pulse allows reflection for the separated beam path interferometer and imposes a `spin echo' effect for the trapped system. (c) The experimental setup. A single photon microwave transition drives internal state transitions in a BEC held in a crossed dipole trap.}


\section{The sensitivity of an atom interferometer}
\label{sect:sensitivity}

In a Ramsey atom interferometer, an input atomic wavepacket is split into two states which are allowed to evolve for a time $T$ until being recombined (see Figure \ref{fig:ramsey}). In this section, we examine the sources of noise in a typical Ramsey interferometer, and present a rigorous analysis of how these affect the sensitivity.

\subsection{Noise sources}
\label{sect:noisesources}

For a temporal Ramsey interferometer (Figure \ref{fig:ramsey}(b)), the noise may be categorised as follows: (i) quantum projection noise, due to the finite number of particles involved in the measurement; (ii) noise introduced by the beam-splitting process; (iii) noise affecting the phase evolution during the evolution time; and (iv) noise on the detection. We discuss each of these in turn.

\subsubsection{Quantum projection noise}
After the final beamsplitter of a $\pi/2-\pi/2$ Ramsey sequence, the atoms exist in a superposition of two states \ket{1} and \ket{2}. Upon measurement, each atom is projected onto either state \ket{1} or state \ket{2}, with probabilities $p_1$ and $p_2$ respectively. Due to the fundamentally random nature of this projection, the final number of atoms in each state will not necessarily equal the expected value, but will fluctuate around this value in any given measurement. This fluctuation is quantum projection noise. Experimentally, the quantity we analyze is the fractional population in state \ket{2}, $p \equiv p_2 = N_2 / N$, where $N=N_1+N_2$. In the absence of squeezing, the projection noise causes binomial fluctuations in this quantity of $\sigma_p = \sqrt{p(1-p)/N}=|\sin{\theta}|/(2\sqrt{N})$. Operating halfway up a fringe, where $\theta=\frac{\pi}{2}$ and the change in population for a given phase shift $\delta p/\delta\phi$ is maximal, the population in each state is $N_1=N_2=N/2$, and the projection noise is $\sigma_p = 1/(2\sqrt{N})$.

Quantum projection noise represents a fundamental limit to the sensitivity of an atom interferometer. The simplest way to reduce this noise is to increase the atom number, since $\sigma_p \propto 1/\sqrt{N}$. This is the motivation for large atom number interferometry; an interferometer made with $10^6$ atoms is intrinsically 30 times more sensitive than one using $10^3$ atoms. As in optical interferometers, the projection noise may also be reduced beyond the limit imposed by quantum projection noise by quadrature squeezing \cite{mckenzie04}. This is discussed further in Section \ref{sect:squeezing}.

\subsubsection{Beam-splitter noise}
In a typical Ramsey interferometer, the first beamsplitter ideally produces an equal superposition $\ket{\psi} = (\ket{1} - i \ket{2}) / \sqrt{2}$ of the two states, known as a $\pi/2$-pulse. The second produces a superposition which depends on the phase difference $\phi$ between the atomic transition and the coupling field accumulated during the evolution time $T$. As the phase of the second beamsplitter is scanned, the population in each state varies sinusoidally, producing Ramsey fringes. The precision of an interferometric measurement is limited by the repeatability of these superpositions \textendash\ any fluctuations in the splitting translates into fluctuations in $p$ at the output of the interferometer. The coupling between the two states can be realised via a single-photon process (for example, microwave coupling between the ground states of \Rb{87}, as in this work) or a two-photon process (for example, Raman coupling using optical transitions to an excited state). In either case, noise in the power or duration of the beamsplitter pulses may limit the sensitivity of the interferometer. The frequency and intensity of the radiation coupling the states must therefore be kept stable, and the energy of the transition itself, which may be affected by external electric or magnetic fields, should not vary.

\subsubsection{Phase evolution noise}
By design, an interferometer is sensitive to differences in the phase accumulated by each state during the interaction time. To measure with high signal-to-noise, the effect being measured (acceleration, rotation, magnetic field, etc.) should exist on a small background of noise. However, any other process which affects the phase of each state differently will contribute to the noise on the output. Factors which affect the entire cloud uniformly, such as fluctuations in magnetic field, result in variations in the final population in each state and lead to an increase in noise. Inhomogeneous effects such as position-dependent shifts due to the trapping potential, local magnetic field inhomogeneity, and shifts due to the collective excitation of the spatial mode cause phase evolution during the time $T$ to proceed at different rates across the trapped cloud. This leads to a classical dephasing and a decrease in fringe contrast.

\subsubsection{Detection noise}
In this and many other recent publications on atom interferometry, the number of atoms in each state is measured using absorption imaging. This technique utilises Beer's Law to extract the atomic column density from a sequence of bright field images. Because the information on atom density is extracted from bright images the classical and quantum noise of the light is coupled into the measurement. With careful experimental design, classical noise can be minimised, leaving only the shot noise of the light field contributing to the readout noise. We analyse this situation in detail in Section \ref{sect:imaging}.

\subsubsection{Other contributions}
Even without performing interferometry, there are a number of experimental difficulties to consider in reliably holding and imaging a large two component atom cloud. We must maintain a non-zero magnetic bias field to suppress spin exchange collisions, and additionally shield the cloud from external RF and microwave fields that may redistribute the interferometric states. The switching of magnetic fields and absorption imaging must not cause measurement errors larger than one part in $10^3$.

\subsection{Sensitivity analysis}

In this section, a general expression is derived for the atomic state after a Ramsey interferometer sequence: identical pulses of arbitrary duration $t$ which bound the free evolution time $T$. The noise in this quantity is then analysed by propagation of uncertainties. In the analysis we make extensive use of the Bloch sphere representation of a two-state quantum system (see Ref. \cite{feynman57}).

\subsubsection{State evolution}

First we define a $\pi/2$ pulse as one which brings an initially vertical Bloch vector into the $x$-$y$ plane, by setting the length of the pulse to
\begin{equation}
t_{\frac{\pi}{2}}=\frac{\cos^{-1}{\left(-\epsilon^{2}\right)}}{\OmegaR}\label{eq:pion2pulselength} \;,
\end{equation}
where $\epsilon=\left|\Delta\right|/\Omega$ is the magnitude of the ratio between the detuning and the Rabi frequency and $\OmegaR=\sqrt{\Omega^{2}+\Delta^{2}}=\Omega\sqrt{1+\epsilon^{2}}$ is the total Rabi frequency \footnote{In this analysis, $0 \leq \epsilon < 1$ since for $\epsilon > 1$ it is not possible to achieve a $50-50$ beamsplitter. The pulse duration required for a $\pi/2$ pulse is shortest on resonance, $t_{\frac{\pi}{2}}\left(\Delta=0\right)=\pi/(2\Omega)$, and approximately equal to this value when $\epsilon\ll 1$.}. 

After the general Ramsey interferometer sequence described above, the final longitudinal spin projection $P_z$, i.e. the $z$-component of the Bloch vector (related to the population in state \ket{2} by $P_z=1-2p$), is given by 
\begin{eqnarray}
P_{z} & = & \alpha^{2}+\left(1-\alpha^{2}\right)\sin\left(\left|\Delta\right|T+\xi\right)\label{eq:Pz3R} \;,
\end{eqnarray}
where 
\begin{eqnarray*}
\alpha = \frac{\epsilon^{2}+\cos\left(\OmegaR t\right)}{\epsilon^{2}+1} \;, \\
\xi = n \pi-\tan^{-1}\left(\frac{\left(1+2\epsilon^{2}\right)\cos\left(\OmegaR t\right)+1}{2\epsilon\sqrt{1+\epsilon^{2}}\sin\left(\OmegaR t\right)}\right) ; \quad n = \cases{1 & if $\quad\sin\left(\OmegaR t\right)<0$\\
0 & if $\quad\sin\left(\OmegaR t\right)\geq0$\\} .\\
\end{eqnarray*}

Using $t=t_{\pi/2}$ from \eqref{eq:pion2pulselength} in the general expression for the final longitudinal spin projection $P_{z}$ after the interferometer sequence, we have 
\begin{equation}
P_{z}(t=t_{\pi/2})=\sin\left[\left|\Delta\right|T-\sin^{-1}\left(1-2\epsilon^{2}\right)\right] \;,
\end{equation}
and so the Ramsey fringe first crosses $P_z = 0$ at the evolution time $T = T_0$ given by
\begin{equation}
\left|\Delta\right|T_{0}=\sin^{-1}\left(1-2\epsilon^{2}\right) \;.
\end{equation}
The time $T_{0}$ is that at which the interferometer is maximally sensitive to relative phase shifts.

\subsubsection{Effect of beamsplitter power fluctuations and resonance fluctuations}
\label{sect:fluctuations}

Suppose the pulse sequence is calibrated for some detuning and Rabi frequency. We seek to find the change in the final longitudinal spin projection for fluctuations in the power $\delta P$ and detuning $\delta\Delta$ of the coupling field, \textit{\emph{while the duration of the pulses $t=t_{\pi/2}$ and evolution time $T=T_{0}$ are kept fixed}}. As such, the change in the spin projection is found by evaluating \footnote{To convert these derived uncertainties in the z-projection of the Bloch vector into uncertainties in the population of state $\left|2\right>$, use $\delta P_{z}= 2 \, \delta p$.}
\begin{eqnarray}
\nonumber (\delta P_{z})^{2} & =\left(\totalD{P_{z}}P\right)^{2}(\delta P)^{2}+\left(\totalD{P_{z}}{\Delta}\right)^{2}(\delta\Delta)^{2}\\
& =\left(\frac{\Omega}{2}\totalD{P_{z}}{\Omega}\right)^{2}\left(\frac{\delta P}{P}\right)^{2}+\left(\totalD{P_{z}}{\Delta}\right)^{2}(\delta\Delta)^{2}\label{eq:noisechainrule}
 \end{eqnarray}
for arbitrary (but constant) $t$ and $T$. Evaluating the expression \eqref{eq:noisechainrule} and then substituting $t=t_{\pi/2}$ and $T=T_{0}$ gives
\begin{eqnarray}
(\delta P_z)^2 = f(\epsilon) \left(\frac{\delta P}{P}\right)^2 + g(\epsilon) \left(\frac{\delta\Delta}{\Omega}\right)^2 \;, \\
\end{eqnarray}
where
\begin{eqnarray}
f(\epsilon) & = &  \left(\frac{\epsilon\left[\cos^{-1}\left(-\epsilon^{2}\right)-\sqrt{1-\epsilon^{4}}\right]}{\left(1+\epsilon^{2}\right)^{3/2}}\right)^{2} \;, \\
g(\epsilon) & = & \left(\frac{\sin^{-1}\left(1-2\epsilon^{2}\right)}{\epsilon}+\frac{2\left[\sqrt{1-\epsilon^{4}}+\epsilon^{2}\cos^{-1}\left(-\epsilon^{2}\right)\right]}{\left(1+\epsilon^{2}\right)^{3/2}}\right)^{2} \;.
\end{eqnarray}
For $\epsilon \ll 1$, this can be approximated as 
\begin{equation}
(\delta P_z)^2 \approx \left(\frac{\pi}{2}-1\right)^{2}\epsilon^{2}\left(\frac{\delta P}{P}\right)^{2}+\left(\frac{\pi}{2\epsilon}\right)^{2}\left(\frac{\delta\Delta}{\Omega}\right)^{2} 
\end{equation}
where the approximation is valid to within $5\units{\%}$ for $\epsilon<0.4$. Each contribution to $\delta P_{z}$ is monotonic in $\epsilon$: fluctuations in $P_{z}$ due to power increase with increasing $\epsilon$ and those due to resonance fluctuation decrease with increasing $\epsilon$. The minimum of $\delta P_{z}$ is achieved when $\rm{d}(\delta P_{z})^{2}/\rm{d}\epsilon=0$ while holding $\delta\Delta/\Omega$ and $\delta P/P$ constant:
\begin{eqnarray*}
\left(\frac{\delta\Delta}{\Omega}\right)^{2}\left(\frac{P}{\delta P}\right)^{2}=-\frac{f'\left(\epsilon\right)}{g'\left(\epsilon\right)} \; \approx \; \epsilon^4 \left(1-\frac{2}{\pi}\right)^2 \;,
\end{eqnarray*}
where the approximation is valid to within $10\units{\%}$ for $ $$\epsilon<0.32$. The optimal detuning can thus be found:
\begin{eqnarray*}
\epsilon_{\rm{opt}} & = & \sqrt{ \left| \frac{\delta\Delta}{\Omega} \frac{P}{\delta P} \right| } \frac{1}{\sqrt{1-2/\pi}}\\
\left| \Delta_{\rm{opt}} \right| & \approx & 1.66 \sqrt{ \left| \delta\Delta \; \Omega \; \frac{P}{\delta P} \right| } \;,
\end{eqnarray*}
which is valid to within $8\units{\%}$ for $\epsilon<0.32$.

The fluctuations in detuning $\delta \Delta$ depend primarily upon fluctuations in the applied magnetic field $B$ and the applied pulse frequency $\omega_{\rm{app}}$. The frequency difference between the $\left|1,0\right\rangle$ and the $\left|2,0\right\rangle$ states in $^{87}\mbox{Rb}$ is found using the Breit-Rabi equation to be $f=f_{0}\sqrt{1+B^{2}x^{2}}$ where $f_{0}\simeq6.834\units{GHz}$ is the hyperfine splitting at $B=0$, and $x = \mu_B(g_J-g_I)/f_{0}$ with $\mu_B$ the Bohr magneton and $g_J$ and $g_I$ the Land\'e g-factors. Thus the fluctuation of the resonant frequency of the atoms will be
\begin{eqnarray*}
\delta\omega_{\rm{res}} & = & 2\pi \, \delta f\\
& = & \frac{2\pi f_{0}Bx^{2}}{\sqrt{1+B^{2}x^{2}}} \, \delta B\\
& = & \kappa\left(B\right)\,\delta B \;,
\end{eqnarray*}
where $\kappa\left(B\right)$ is defined in the last line. The fluctuation in the detuning $\Delta=\omega_{\rm{res}}-\omega_{\rm{app}}$ is then given by
\begin{eqnarray*}
\left( \delta \Delta \right)^{2} & = & \left(\delta\omega_{\rm{res}}\right)^{2}+\left(\delta\omega_{\rm{app}}\right)^{2}\\
& = & \kappa^{2}\left(\delta B\right)^{2}+\left(\delta\omega_{\rm{app}}\right)^{2}\,.
\end{eqnarray*}

To minimise sensitivity to power fluctuations, a small detuning $\Delta$ during the $\pi/2$-pulses is required. However, this necessitates a longer evolution time for the spin to precess by $\sim \pi/2$, resulting in more sensitivity to resonance fluctuations. Using a small (ideally zero) detuning during the pulses and a different detuning $\Delta_1$ during the evolution time \footnote{This can be achieved by changing the atomic resonance by e.g. applying a magnetic field, or by changing the frequency of the reference oscillator.} results in minimal sensitivity to fluctuations. Repeating the above analysis for $\Delta = 0$ then gives
\begin{equation}
(\delta P_{z})^2 = \left(\frac{\pi}{4}\right)^4 \left(\frac{\delta P}{P}\right)^4 + 4\left(\frac{\delta\Delta}{\Omega}\right)^2 + \frac{\pi^2}{4} \left(\frac{\delta\Delta_1}{\Delta_1}\right)^2 \;,
\end{equation}
where the second order power fluctuations have been included. In this way $P_z$ can be made insensitive to power fluctuations to first order.


\section{An optically trapped atom interferometer}
\label{sect:interferometer}

In this section we present our Ramsey trapped atom interferometer operating on the $\hf{F=1}{m_F=0} \rightarrow \hf{F=2}{m_F=0}$ transition in Bose-condensed \Rb{87}.

\subsection{Apparatus}

Our experimental apparatus for producing optically trapped BECs of \Rb{87} and \Rb{85} is described in detail elsewhere \cite{altin10}. In brief, we first load a retro-reflected 3D magneto-optical trap (MOT) with $10^{10}$ \Rb{87} atoms in $5$ seconds from a 2D-MOT source. The atoms undergo polarization-gradient cooling for $20\units{ms}$ before being loaded into a quadrupole magnetic trap (100\units{G/cm} field gradient, $1\units{G}=10^{-4}\units{T}$) in the \hf{F=1}{m_F=-1} state. The cloud is transported over $4\units{cm}$ to a quadrupole-Ioffe configuration (QUIC) magnetic trap with a bias field of $B_0=3.4\units{G}$ and trapping frequencies $\omega_z = 2\pi\times16\units{Hz}$ and $\omega_\rho = 2\pi\times156\units{Hz}$. Radiofrequency forced evaporation then reduces the temperature of the sample to $10\units{$\mu$K}$ over 15\units{s}. A crossed-beam dipole trap from a $1090\units{nm}$, $20\units{W}$ fibre laser is then abruptly switched on and a homogeneous bias magnetic field of $160\units{G}$ is added along the $z$-axis. One beam of the dipole trap propagates along the weak axis of the QUIC trap with a waist of $150\units{$\mu$m}$, while the other is almost perpendicular and in the horizontal plane with a waist of $200\units{$\mu$m}$. Approximately $10^7$ atoms are thus held in a hybrid trap in which the radial confinement is provided by the dipole trap and axial confinement by the magnetic field curvature of the QUIC trap.

Here the atoms are cooled for a further $7\units{s}$ by lowering the intensity of the dipole trap laser. During this time the QUIC magnetic field is also ramped to zero leaving the atoms in a pure crossed dipole trap ($\omega_{x,y,z} = 2\pi\times(50,57,28)\units{Hz}$) with a homogeneous bias field of $160\units{G}$. Finally, the atoms are transferred to the $m_F=0$ state via a $5\units{ms}$ Landau-Zener rf sweep. The result is an optically confined Bose-Einstein condensate containing $2\times10^6$ \hf{F=1}{m_F=0} atoms. For an interferometry run, the bias field can be set to any value between $0\units{G}$ and $200\units{G}$. In this work we ramp the bias field down to $4\units{G}$.

We create a trapped Ramsey interferometer by coupling the \hf{F=1}{m_F=0} and \hf{F=2}{m_F=0} hyperfine states with a single microwave photon at $6.835\units{GHz}$. The microwave signal is produced by a pulse-gated Rohde \& Schwarz SMR20 microwave generator locked to a rubidium frequency standard (SRS FS725). The microwaves are delivered to the experiment via a simple quarter-wave dipole antenna, essentially a $1\units{cm}$ piece of copper wire soldered to an SMA connector. The antenna is glued in place as close as possible to the atom cloud. With this setup, we are able to couple the clock states with maximum Rabi frequencies of around $5\units{kHz}$. 

A typical interferometry sequence is as follows. A $300\units{$\mu$s}$ $\pi/2$ pulse creates an equal superposition of the \hf{F=1}{m_F=0} and \hf{F=2}{m_F=0} states (hereafter referred to as the \ket{1} and \ket{2} states respectively). After an evolution time $T$, another $\pi/2$ pulse is applied and the trap is switched off, allowing the atom cloud to expand freely. During this expansion, a strong magnetic field gradient is applied to spatially separate the states \ket{1} and \ket{2} using the second-order Zeeman shift. After $20\units{ms}$ of expansion, a pulse of light resonant with the $\ket{F=1} \rightarrow \ket{F^\prime=2}$ transition is applied to optically pump the \ket{1} atoms out of the lower ground state, before both states are imaged by absorption on the $\hf{F=2}{m_F=2} \rightarrow \hf{F^\prime=3}{m_{F^\prime}=3}$ transition.

Figure \ref{fig:visibility}(a) shows Ramsey fringes in the fractional population of the \ket{2} state after the second $\pi/2$ pulse, recorded by scanning the detuning of the microwave pulses from the atomic resonance. Fringes are shown for several different values of the evolution time $T$. Measuring only the relative population $p_2 = N_2 / (N_1 + N_2)$ allows run-to-run variations in total atom number to be normalized out. The total number of atoms in these measurements is $N_1 + N_2 = (1.0 \pm 0.1)\times10^6$.

\fig{.6\textwidth}{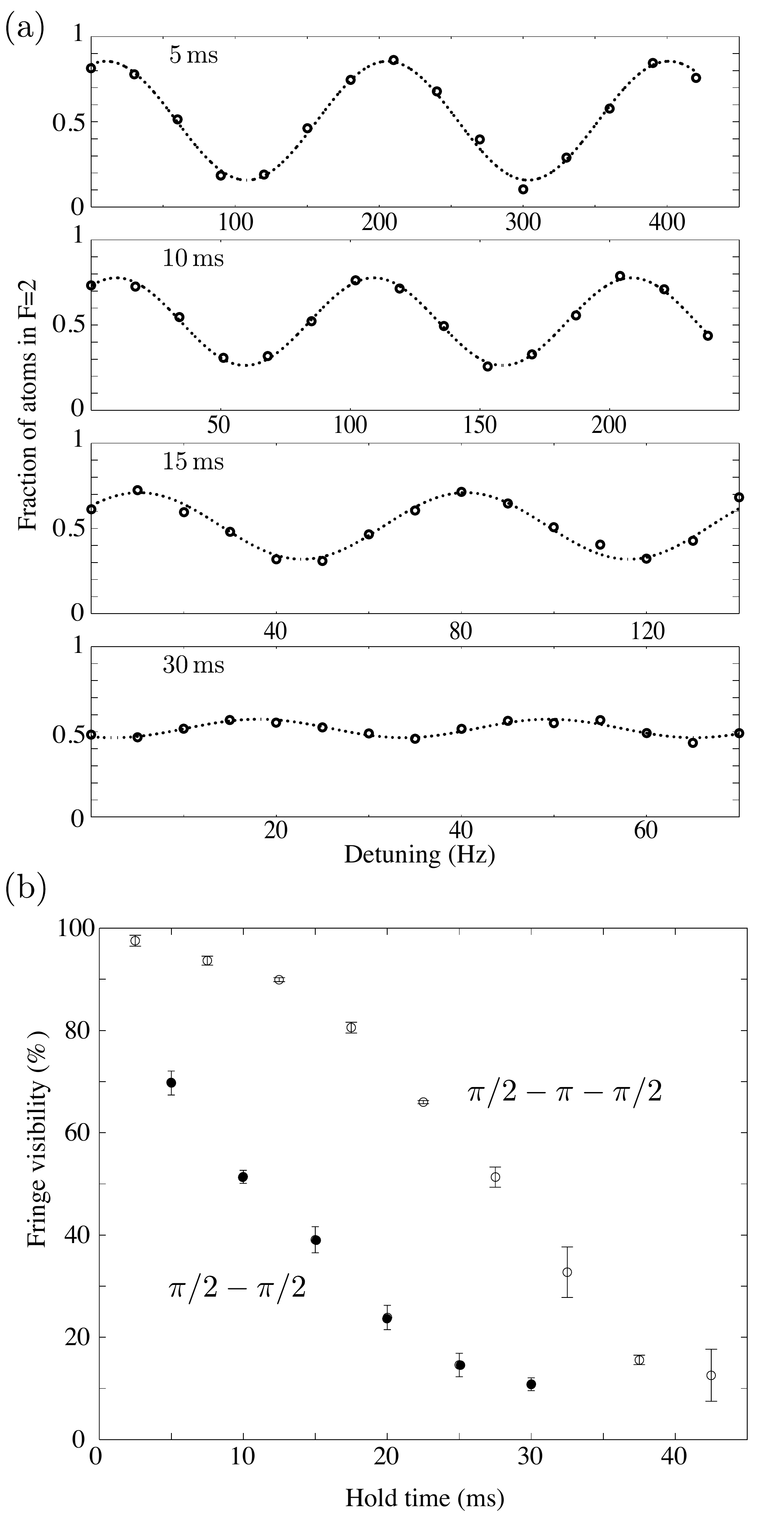}{$(a)$ Decreasing fringe visibility with increasing evolution time in a $10^6$ atom interferometer. The uncertainties are smaller than the data points. $(b)$ Fringe visibility with ($\pi/2 - \pi - \pi/2$) and without ($\pi/2 - \pi/2$) a `spin echo' $\pi$ pulse. Error bars represent primarily statistical uncertainties, systematic effects are dominated by number fluctuations which are normalized out in the measurement of relative number.}

To assess the performance of our system relative to the shot noise limit, we operate the interferometer at mid-fringe with $N=5\times10^5$ atoms and an evolution time of $5\units{ms}$ with no spin-echo pulse. We find that the run-to-run standard deviation is a factor of 15 higher than the projection noise limit for this atom number, corresponding to a fluctuation of $2.1\units{\units{\%}}$ in the splitting probability $p_2$ and a measured interferometric sensitivity of $1.8\units{mrad}$ over 30 minutes.

\subsection{Fringe visibility}

For short evolution times $T \lesssim 10\units{ms}$ and total atom numbers $\sim10^6$, we observe high contrast interference fringes with a visibility approaching $100\units{\%}$ (Figure \ref{fig:visibility}(a)). As $T$ is increased to $30\units{ms}$, however, the visibility decays to below $20\units{\%}$. Some of this loss of coherence is due to deterministic effects such as inhomogeneous broadening of the transition frequency across the trapped cloud. It is possible to reverse this dephasing by applying a `spin-echo' $\pi$ pulse between the two Ramsey pulses (Figure \ref{fig:ramsey}(b)). This has the effect of flipping the state vectors on the Bloch sphere, thus reversing their spread during the second half of the evolution time. Figure \ref{fig:visibility}(b) shows the effect on fringe visibility of adding a spin-echo pulse to an interferometer with $N = 1\times10^6$ atoms. The pulse clearly increases the coherence time, with the time taken for the fringe visibility to drop to $70\units{\%}$ increasing from $5\units{ms}$ to $20\units{ms}$.

We also contend with irreversible inelastic losses from the interferometer states. The magnetically sensitive states \hf{F=1}{m_F=\pm1}, \hf{F=2}{m_F=\pm1} can be populated by spin-exhange collisions \cite{klempt10}, as shown in Figure \ref{fig:spinexchange}(a). The inelastic loss rates also differ between the two interferometer states. Figure \ref{fig:spinexchange}(b) shows the populations in each state over 1\units{s} in the optical trap. The decay of state \ket{2} is markedly faster than of state \ket{1} due to inelastic dipolar relaxation from the upper hyperfine level. These losses contribute to the loss in fringe visibility as the evolution time is increased, but are also compensated for by a spin echo pulse. This is because the assymetric losses result in the Bloch vector falling out of the $x$-$y$ plane during the evolution time due to heavier loss from state \ket{2}. The echo acts to bring it above the $x$-$y$ plane, where it then falls back into the plane prior to the second $\pi/2$ pulse.

\fig{0.7\textwidth}{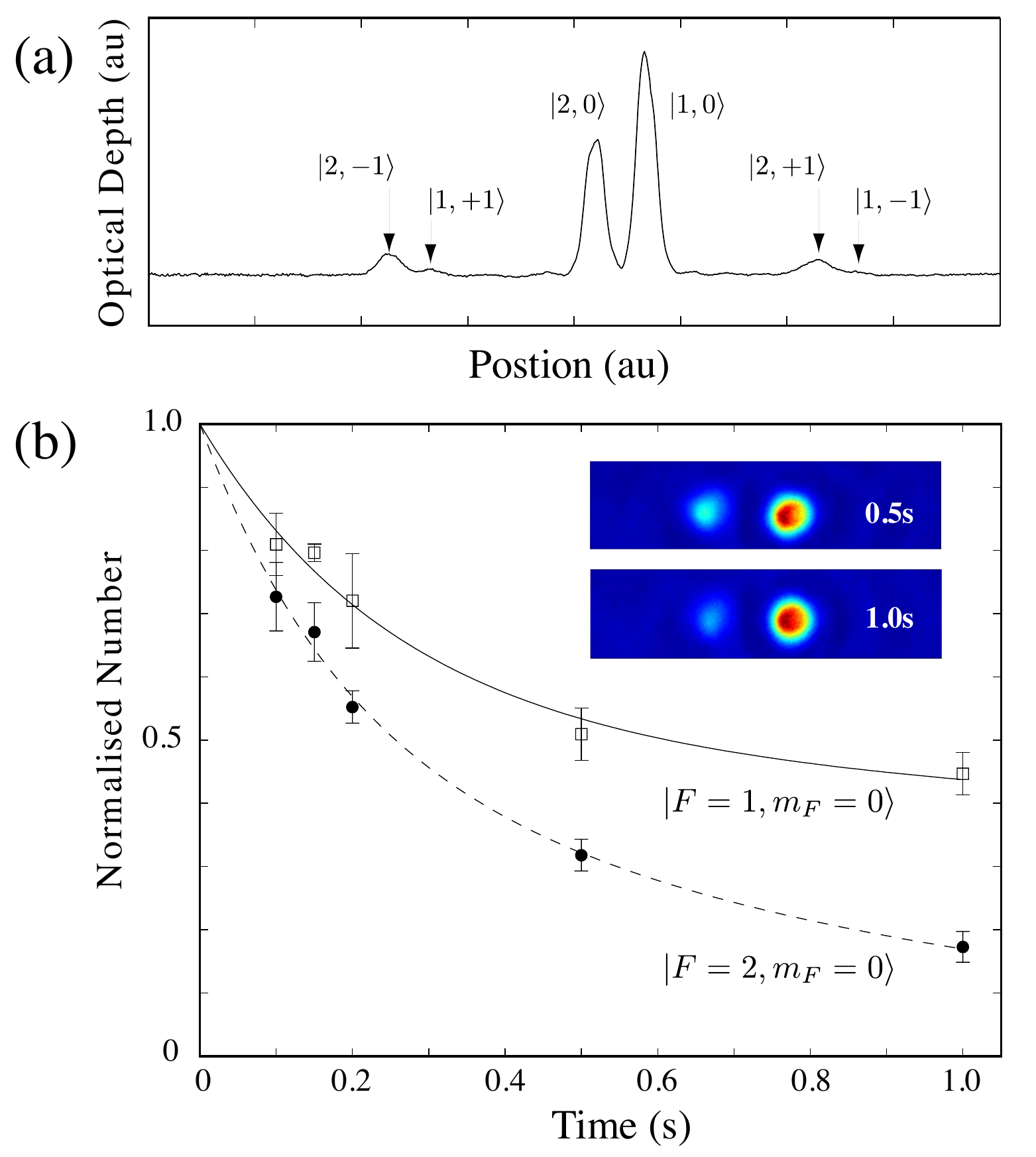}{Inelastic processes which lead to a loss in fringe contrast. $(a)$ Integrated optical depth of a sample of $N=(5.0\pm0.2)\times10^5$ atoms. The \hf{F=1}{m_F=\pm1} and \hf{F=2}{m_F=\pm1} states are populated by spin-exchange collisions. $(b)$ Measured loss rates for the two interferometric states. The inset shows the interferometer states \ket{1} and \ket{2} after 0.5\units{s} and 1\units{s} hold times. The error bars represent the combined statistical and systematic uncertainties.}

The remaining fringe decay is a combination of technical noise and irreversible dephasing due to atomic interactions \cite{horikoshi06}. One would expect the majority of technical noise (e.g. current noise in the magnetic field coil power supplies) to affect the entire cloud uniformly, which would cause an increase in the noise on the fringes but no decrease in visibility. To a good approximation, any remaining inhomogeneous technical noise should cause a decay in fringe visibility independent of atom number, while interaction-induced dephasing will affect large atom number condensates more due to their higher density (this will be discussed in detail below). For an interferometer with $N=5\times10^4$ atoms, we measure a decoherence time of over $800\units{ms}$, with the fringe visibility still at $80\units{\%}$ after $200\units{ms}$. Recent work has also shown that a \emph{self-rephasing} effect can lead to coherence times of many tens of seconds \cite{deutsch10}. We have not yet observed this effect in our lab. 

Thus we conclude that, in our setup, interaction-induced dephasing is the limiting factor for achieving long coherence times with large atom numbers. This could be overcome by making the samples more dilute (using a weaker trap) or by reducing the interactions between atoms using a Feshbach resonance. Manipulating the scattering length has been demonstrated to give a large increase in interferometric contrast in Bloch oscillations \cite{gustavsson08,fattori08}.

\subsubsection{Effect of interactions}

To see how the interactions affect the fringe contrast in a simple $\pi/2- \pi/2$ interferometer scheme, consider a two mode semiclassical model of the BEC system \cite{dalfovo99}, where we ignore the dynamics of the spatial mode. In our interferometer the condensate is split into two spatially overlapping hyperfine states. Assuming there are no multimode excitations, the mean field wavefunction may be written as
\begin{equation}
\Psi(\mathbf{r},t) = c_1(t)\psi_1(\mathbf{r})|1\rangle + c_2(t)\psi_2(\mathbf{r})|2\rangle \;,
\end{equation}
where $\psi_1(\mathbf{r})$ and $\psi_2(\mathbf{r})$ represent the normalised spatial mode functions of the the $|1\rangle$ and $|2\rangle$ components of the condensate. The number of atoms in each mode is given by $|c_1|^2=N_1$, $|c_2|^2 = N_2$, with $N_1+N_2=N$, and $\int |\Psi|^2\, d^3\mathbf{r}=N$, normalized to the total number of atoms.  From the Gross-Pitaevski equation, we obtain the equations of motion for the coefficients $c_1$ and $c_2$:
\numparts
\begin{eqnarray}
i \frac{dc_1}{dt} = \omega_1 c_1 + (g_{11} N_1 + g_{12} N_2)c_1 \equiv \omega_a c_1\\
i \frac{dc_2}{dt} = \omega_2 c_2 + (g_{22} N_2 + g_{12} N_1)c_2 \equiv \omega_b c_2 \;,
\end{eqnarray}
\endnumparts
where $\hbar\omega_{1,2}$ are the linear contributions to the energy of each state, and 
\begin{equation}
g_{ij} = \frac{U_{ij}}{\hbar}\int  |\psi_i(\mathbf{r})|^2 |\psi_j(\mathbf{r})|^2\, d^3\mathbf{r} \, .
\end{equation}
For large Bose-Einstein condensates, we can use the Thomas-Fermi approximation for $\psi_i$:
\begin{equation}
|\psi_i(\mathbf{r})|^2 = \frac{1}{N}\frac{\mu - V(\mathbf{r})}{U_{11}} \;,
\end{equation}
where $N$ is the total number of atoms, $\mu$ is the chemical potential, and $U_{11}$ and $V(\mathbf{r})$ denote the $s$-wave interaction constant and the external potential, respectively, for the $|1\rangle$ component. This gives
\begin{equation}
g_{ij}= \frac{U_{ij}}{U_{11}} \frac{2^{1/5}}{7\hbar} \left(\frac{15 U_{11}}{\pi}\right)^{2/5} \left(\frac{m\bar\omega^2}{N}\right)^{3/5} \,,
\end{equation}
where $\bar{\omega}$ is the geometric mean of the trapping frequencies for each dimension. It should be noted that $g_{ij}$ is a function of the total number of atoms $N$, but not the number of atoms in each mode. 

The rate at which the relative phase between the two modes evolves is
\begin{eqnarray}
\frac{d \phi}{dt} &=& \omega_a - \omega_b \nonumber \\
&=& (\omega_1-\omega_2) + (g_{11}-g_{12})N_1 - (g_{22}-g_{12})N_2 \;. \label{eq_freq_diff}
\end{eqnarray}

Up to this point the evolution of the relative phase is completely deterministic, and hence reversible. However, we now take into account that when the atoms are coupled by the beam splitter, there is an uncertainty in the number in each state. Assuming a 50/50 beam splitter, we can represent this number uncertainty by setting $N_1 = N/2 + \Delta N$, and since the total number of atoms is a conserved quantity, $N_2 = N/2 - \Delta N$. Inserting this into \eqref{eq_freq_diff} gives
\begin{equation}
\frac{d \phi}{dt} = (\omega_1-\omega_2) + (g_{11}-g_{12}) \left(\frac{N}{2} + \Delta N\right) - (g_{22} -g_{12}) \left(\frac{N}{2} - \Delta N\right) \;.
\end{equation}
As fluctuations in number due to spin projection noise are binomial (Section \ref{sect:noisesources}), this uncertainty in the number difference will be $\Delta N \approx\sqrt{N}/2$. Ignoring the deterministic part of the phase evolution, the uncertainty in the evolution rate is
\begin{equation}
\Delta ( \omega_a - \omega_b) = \frac{1}{2}(g_{11} - 2g_{12} + g_{22})\sqrt{N} \;,
\end{equation}
and so the phase uncertainty after time $T$ is
\begin{equation}
\Delta \phi = \Delta ( \omega_a - \omega_b) T = \frac{1}{2}(g_{11} - 2g_{12} + g_{22}) T \sqrt{N} \;.
\end{equation}
A similar analysis shows that fluctuations in the total number will lead to an additional phase diffusion of $\Delta\phi = (g_{11}-g_{22}) T \Delta N$, where $\Delta N$ represents shot-to-shot fluctuations in the total number. 

Now, if we consider a spin echo interferometer ($\pi/2-\pi-\pi/2$), the phase diffusion due to fluctuations in the total number is removed, since the phase difference accumulated before the spin echo pulse is cancelled by that accumulated after it. However, this does not apply to the fluctuations in the relative number difference, as can be shown by tracking the phase of $c_1$ and $c_2$. Before the $\pi$ pulse, after evolving for a time $t_1 = \Delta t$, the phase of each mode is
\numparts
\begin{eqnarray}
\phi_1(t_1) &=& \left(\omega_1 + g_{11} \left(\frac{N}{2} +\Delta N\right) + g_{12} \left(\frac{N}{2} -\Delta N\right)\right)\Delta t \\
\phi_2(t_1) &=& \left(\omega_2 + g_{22} \left(\frac{N}{2} - \Delta N\right) + g_{12} \left(\frac{N}{2} +\Delta N\right)\right)\Delta t \;.
\end{eqnarray}
\endnumparts
The spin echo pulse does two things: it reverses the populations such that the population of mode 2 becomes the population of mode 1, and it also performs the mapping $c_1 \rightarrow ic_2$, $c_2 \rightarrow ic_1$. So, directly after the spin echo pulse, at time $t_2$, we have
\numparts
\begin{eqnarray}
\phi_1(t_2) &=& \phi_2(t_1) + \frac{\pi}{2} \\
\phi_2(t_2) &=& \phi_1(t_1) + \frac{\pi}{2} \;.
\end{eqnarray}
\endnumparts
Evolving by another time $\Delta t$, the phase of each mode becomes
\numparts
\begin{eqnarray}
\phi_1(t_3) &=& \phi_1(t_2) + \left(\omega_1 + g_{11} \left(\frac{N}{2} - \Delta N \right) + g_{12} \left(\frac{N}{2} +\Delta N\right)\right)\Delta t \\
\phi_2(t_3) &=& \phi_2(t_2) + \left(\omega_2 + g_{22} \left(\frac{N}{2} + \Delta N \right) + g_{12} \left(\frac{N}{2} -\Delta N\right)\right)\Delta t  \;,
\end{eqnarray}
\endnumparts
where we note that the populations have been reversed, i.e. the population in mode 1 is now $N/2 - \Delta N$. Taking the difference in the phases at $t = t_3$ gives
\begin{eqnarray}
\Delta\phi &=& \phi_2(t_3) - \phi_1(t_3) \nonumber \\
&=& 2(g_{11}-2g_{12}+ g_{22})\Delta N \Delta t \nonumber \\
&=& (g_{11}-2g_{12}+ g_{22})\sqrt{N}\Delta t \;,
\end{eqnarray}
or, in the Thomas-Fermi limit,
\begin{equation}
\Delta\phi = \frac{(a_{11} - 2 a_{12} + a_{22})}{a_{11}}\frac{2\left(15 \sqrt{m} \hbar^2 \bar{\omega}^3 a_{11}\right)^{2/5}}{7\hbar} \frac{\Delta t}{N^{1/10}} \, ,
\end{equation}
where $a_{ij}$ is the $s$-wave scattering length for collisions between state $i$ and $j$. 
Thus, the phase diffusion arising from interactions and the uncertainty in the \emph{number difference} cannot be removed with a spin echo, leading to an irreversible phase evolution in the system. The phase diffusion due to fluctuations in the \emph{total number} can be removed by this technique. For the parameters used in this experiment, we calculate the phase diffusion rate to be $\approx 50$ mrad s$^{-1}$. However, it should be noted that this rate is small due to the favourable scattering properties of $^{87}$Rb, since $(a_{11} -2 a_{12} + a_{22})/a_{11} \approx -0.02$.

The interactions in the condensate also cause an overall density dependent shift to the resonance frequency \cite{gorlitz03}. Here, we do not consider this effect to be a source of noise but rather a measurable systematic effect.

\subsection{Miscibility}
\label{section:miscibility}

The intra- and inter-species scattering lengths in the $\hf{F=1}{m_F=0}$ and $\hf{F=2}{m_F=0}$ states are predicted by full coupled channel simulations \cite{chin10} using the potentials of \cite{strauss10} to be $a_{11} = 100.9\,a_0$, $a_{12} = 98.9\,a_0$, and $a_{22} = 94.9\,a_0$, where $a_0$ is the Bohr radius. The miscibility parameter $\mu=a_{11}a_{22}/a_{12}^2 \sim 0.98 < 1$ predicts that the states will be immiscible, since the cross-species repulsion exceeds the self-repulsion. Spatial dynamics driven by differences in scattering length are clearly observed in the magnetic states $\hf{F=1}{m_F=-1}$ and $\hf{F=2}{m_F=1}$ after $T\sim 10\units{ms}$, which causes a loss of interferometric contrast \cite{anderson09}. However, we do not observe any dynamics of the $\hf{F=1}{m_F=0}$ and $\hf{F=2}{m_F=0}$ states after up to $1\units{s}$ evolution times (see inset of Figure \ref{fig:spinexchange}(b)), despite being able to detect domain separation in other states. This suggests either that spatial dynamics are occurring (and contributing to the decay in fringe visibility as $T$ is increased) but are corrupted by our imaging procedure and thus not observed, or that the scattering lengths in our system are different to the theoretically predicted values. If the clock states are indeed miscible, this would make them more favourable for interferometry since miscibility minimises mean field driven dephasing and allows good wavefunction overlap for the recombination pulse.

\subsection{Beam-splitter stability}
\label{sect:stability}

For small detuning $\Delta$, the fluctuations in $P_z$ after a single $\pi/2$ beamsplitter are $\delta P_{z} \approx (2-\pi/2) (\Delta/\Omega) \, \delta\Delta/ \Omega + (\pi/4) \, \delta P/P$. Operating on resonance thus makes the beamsplitter immune to all relative frequency fluctuations to first order and requires the minimum rf power for a given $\pi/2$ pulse time. This gives good suppression of frequency instabilities in the microwave source and fluctuations in the resonance frequency due to changes in the magnetic bias field. As a rough estimate, a relative transition frequency fluctuation between runs of $10\units{Hz}$, which could be caused by a $10\units{Hz}$ uncertainty in the microwave frequency or a $2\units{mG}$ uncertainty in the $4\units{G}$ magnetic bias field, would cause only a $0.006\units{\%}$ fluctuation in transition probability on resonance for a 50/50 beam-splitter. This increases to a $0.1\units{\%}$ fluctuation $100\units{Hz}$ off-resonance. The effect of magnetic field noise is reduced further at lower fields, since the two $m_F = 0$ states are sensitive only to second order in the magnetic field. Our measured background fluctuations are $4\units{mG}$ peak-to peak, predominantly at 50\units{Hz}. Our home built bias field power supply has a measured rms current noise of 1 part in $10^{6}$. 

\fig{.75\textwidth}{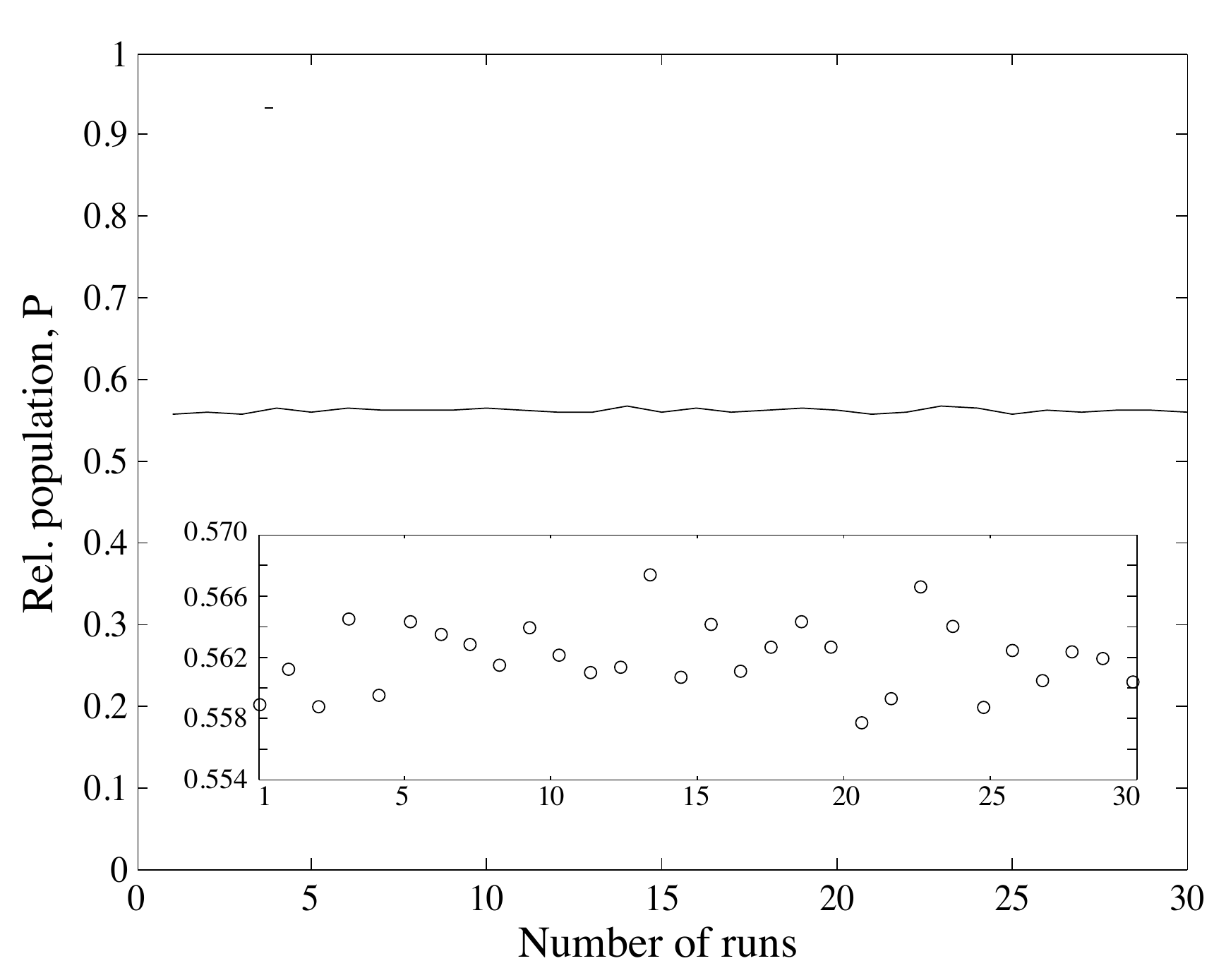}{Performance of a single beamsplitter, showing the relative population as a function of the number of runs. This data has fluctuations a factor of 2.2 above atom shot noise on the $3\times10^5$ atoms used.}

While frequency fluctuations may be suppressed by operating on resonance, the amplitude of the microwave pulse affects the transition probability on a single beam-splitter linearly. We have measured the power variations in our $300\units{$\mu$s}$ microwave $\pi/2$ pulses to be approximately $0.5\units{\%}$, which precludes the observation of atom shot noise from a single beamsplitter with more than $10^4$ atoms. Indeed, we observe fluctuations a factor of 2.2 above atom shot noise in a $\pi/2$ pulse with $3\times10^5$ atoms (see Figure \ref{fig:bsnoise}).

When operated on resonance, however, the spin projection $P_z$ after a full interferometer sequence is insensitive to power fluctuations to first order (see Section \ref{sect:fluctuations}). Thus the $0.5\units{\%}$ power variation translates to only a $\sim 0.0015\units{\%}$ variation on the interferometer fringes, which is a factor of $30$ below the projection noise limit on our $10^6$ atom condensates.


\section{Shot-noise limited absorption imaging}
\label{sect:imaging}

Absorption imaging is one of the most commonly used methods of studying ultra-cold atom clouds, and is our primary tool for extracting information from our experiments. In order to directly observe squeezing and a consequent increase in the sensitivity of an interferometer, it is necessary that the noise introduced by the imaging system be well below the atomic shot noise fluctuations. It has been shown that using CCD cameras with high quantum efficiency (QE), it is possible to directly measure atom number fluctuations at and below the projection noise limit on small ensembles in a series of absorption images \cite{greiner05}. Here we investigate whether this is possible for large clouds. We show that using a CCD camera with a quantum efficiency of only $17\units{\%}$ it is possible to achieve a signal-to-noise ratio of nearly $10^4$, almost an order of magnitude above that required to observe atomic shot noise on a $10^6$ atom condensate. This is enough to observe a $9\units{dB}$ reduction in atomic shot noise.

Absorption imaging of cold clouds is performed by analysing two consecutive CCD images, one containing a shadow of the cloud and the other a bright-field image in the absence of the atom cloud. We denote the number of electrons in the array of wells on the camera by $e_i$ and $e_f$. These two images are processed pixel-wise across the array to find the number of atoms $N_{px}$ imaged at each pixel by the following equation:
\begin{equation}
N_{px} = c_0 \left( L \, \ln\frac{e_i}{e_f}+ \frac{e_i-e_f}{e_{sat}} \right) \;,
\label{eqn:absnumber}
\end{equation}
where $c_0 = 2\pi P^2 / (3\lambda^2 M^2)$ is the resonant absorption cross-section scaled by the pixel area $P$ and magnification $M$, $e_{sat}$ is the electron count corresponding to the saturation intensity $I_{sat} = 1.67\units{mW/cm$^2$}$, and $L = (4\Delta^2+\Gamma^2 )/\Gamma^2$ accounts for a detuning $\Delta$ of the imaging light from the $\hf{F=2}{m_F=2} \rightarrow \hf{F^\prime=3}{m_{F^\prime}=3}$ transition, where $\Gamma = 2\pi\times6.067\units{MHz}$ is the natural linewidth. The relationship between the electron counts recorded at a pixel and the intensity $I_{px}$ incident on that pixel is $e_{px} = \eta P \tau I_{px}/ (\hbar\omega)$, where $\eta$ is the camera quantum efficiency, $\tau$ is the exposure time, and $\hbar\omega$ is the energy per photon. The light intensity at the position of the atoms is further scaled by the magnification of the imaging system.

Assuming that all classical noise can be removed from the image, either via careful experimental setup or by numerical post-processing, there are only two noise sources in an image of the interferometric states: projection noise on the relevant number and photon shot noise, which manifests as fluctuations in $e_{px}$. Read noise, dark counts and digitisation error are all negligible in our setup. The atom number in each cloud fluctuates according to the projection noise $\sigma_a = \sqrt{Np\left(1-p\right)}$, assuming a coherent spin state. The electron number fluctuations are Poissonian if the laser light incident on the camera is Poissonian, and so $\sigma_{el} = \sqrt{e_{px}}$. This can be expressed as a fluctuation $\sigma_{det}$ in measured atom number of
\begin{equation}
\sigma_{det} = c_0 \sqrt{\sum_x \left[ e_{i,x} \left( \frac{1}{e_{sat}} + \frac{L}{e_{f,x}} \right)^2 + e_{f,x} \left( \frac{1}{e_{sat}} + \frac{L}{e_{i,x}} \right)^2 \right]} \;.
\end{equation} 
where the sum is taken over all pixels $x$ in a region of interest. The total noise in the image is then $\sigma = \sqrt{\sigma_a^2 + \sigma_{det}^2}$. To directly observe atom number fluctuations, the detector noise $\sigma_{det}$ must be well below the atom shot noise $\sigma_a$.

The quantity $\sigma_{det}$ is completely determined by the parameters of the imaging system. It is therefore possible to predict in advance the best way to image a cloud of a certain size and number. We have performed numerical simulations of absorption images, by solving \eqref{eqn:absnumber} for $e_f$, to find the imaging parameters which minimise $\sigma_{det}$. The simulation determines the number of electrons accumulated in each pixel of the CCD array during an absorption image of a typical condensate following ballistic expansion. Figure \ref{fig:imaging} plots the cumulative standard deviation as a function of the number of runs for an imaging system optimised to image a single Bose-Einstein condensate of $N = 10^6$ atoms released from our crossed dipole trap. With the atom number fluctuations $\sigma_a$ set to zero, we find the lowest achievable noise in the images to be $\sigma = 113$ atoms, implying a detection noise of $\sigma_{det}/N = 0.0113\units{\%}$. The imaging parameters are: intensity $15 I_{sat}$ on resonance, magnification 8, exposure time $100\units{$\mu$s}$, and expansion time $30\units{ms}$. Notably, the simulation assumes a camera with a low QE of $17\units{\%}$, which can be a factor of twenty lower in cost compared to the high QE cameras used in previous work \cite{greiner05}. This level of detector noise would permit observation of atom shot noise on a $10^6$ atom condensate with a signal-to-noise ratio of 9, or an atomic squeezing-enhanced interferometric measurement with sensitivity $9\units{dB}$ higher than the projection limit. For comparison, with the same camera, an $N = 10^5$ condensate can be imaged with a ratio of 5 between atom and photon shot-noise, with a magnification of 2 at saturation intensity. This is the configuration used in the current work. 

\fig{0.75\textwidth}{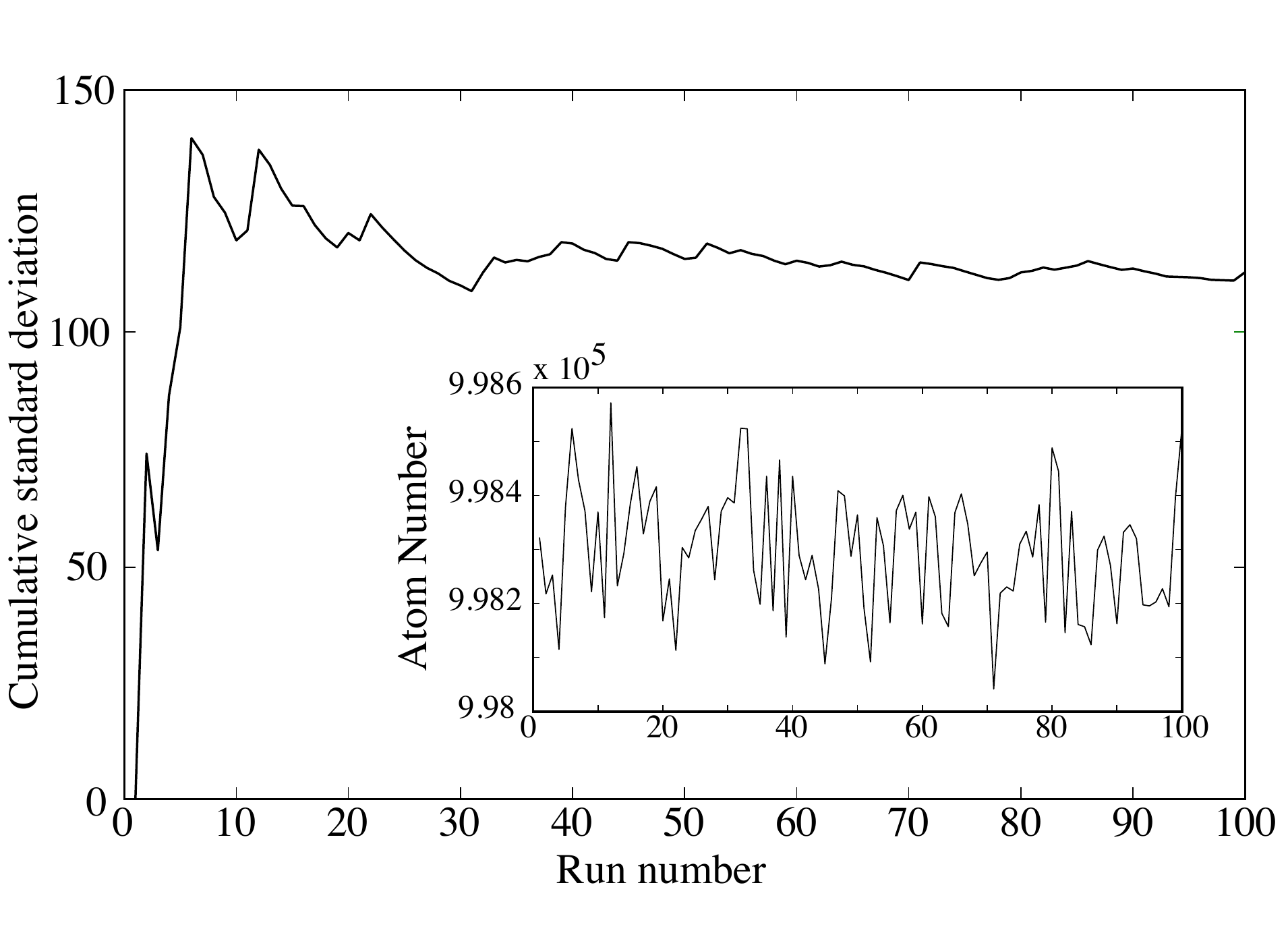}{Simulated detector shot noise, showing the cumulative standard deviation and in the inset the actual fluctuation in measured atom number from run to run in the simulation.}

In practice, there are several further considerations when attempting to observe atomic shot noise by absorption imaging. Firstly, it is imperative that all sources of classical noise be removed from the imaging system. In our experience, the required stability can be achieved by careful design (removing unnecessary optical elements, using high stability mounts, isolating the beam path from air currents, removing dust particles, etc.). Secondly, the \emph{absolute} atom number must be accurately calibrated by an independent method. There are several ways to do this, including analysing the phase transition from thermal cloud to BEC, or measuring the atom shot noise as a function of number. However, the atom number need not be calibrated to the level of the shot noise fluctuations; typically the number will be known in absolute terms to the $2-3\units{\%}$ level, giving negligible error on the predicted shot noise. Finally, the number of experimental runs performed must be sufficient for the measured standard deviation in the atom number to be an accurate representation of its asymptotic value. As can be seen in Figure \ref{fig:imaging}, $20-30$ runs is typically sufficient to have an excellent estimation of the asymptotic variance.


\section{Interferometry beyond the standard quantum limit}
\label{sect:squeezing}

Recent theoretical schemes to produce squeezing in an atom laser \cite{johnsson07}, as well as the experimental demonstration of squeezing-enhanced sensitivity of trapped atom interferometry \cite{riedel10,gross10}, point the way to future advances in precision measurement. We will now consider the enhancement in interferometer sensitivity by using a simple on-axis twisting scheme. In this scheme, a double interferometer ($\pi/2-\pi/2-\pi/2$) is used. The first interferometer prepares the quantum squeezing, and the second uses the squeezed state to perform a measurement with a sensitivity greater than the projection noise limit. To investigate how this would enhance the sensitivity of our setup, we have used a simple theoretical model, which is described in \cite{haine09}. The sensitivity in our estimate of the phase shift in the measurement interferometer is fundamentally limited by the projection noise in the number difference measurement. The uncertainty in the phase due to this effect is
\begin{equation}
\Delta \phi = \frac{\sigma_{P_z}}{|\frac{d \langle P_z\rangle}{d\phi}|} \;,
\end{equation}
where as before $N_{total} P_z \equiv N_1-N_2$ at the interferometer output. For uncorrelated atoms, $\Delta \phi = 1/\sqrt{N_{total}}$ \cite{dowling98}.

Figure \ref{fig:squeezing} shows the phase sensitivity as a function of the phase of the second interferometer. The quantum projection noise limit is shown for reference. The case shown gives greater than a factor of $2$ enhancement in the phase sensitivity, which is equivalent to increasing the atom number by a factor of $4$. As the effective interaction strength for \Rb{87} is quite small, it takes a relatively long time ($20\units{ms}$) to prepare this level of squeezing. The maximum squeezing achievable for this system occurs after $\sim 50$ ms. However, we considered this evolution time unrealistic as effects such as particle loss and technical noise will start to become relevant over this time scale. If we employed a Feshbach resonance to enhance the effective interaction strength, it would be possible to achieve a phase sensitivity of $\Delta \phi \times \sqrt{N} = 0.16$ with a $0.25\units{ms}$ evolution time. Any further level of quantum squeezing would not give significantly further enhancement, as we would then become limited by the detection noise.

\fig{0.75\textwidth}{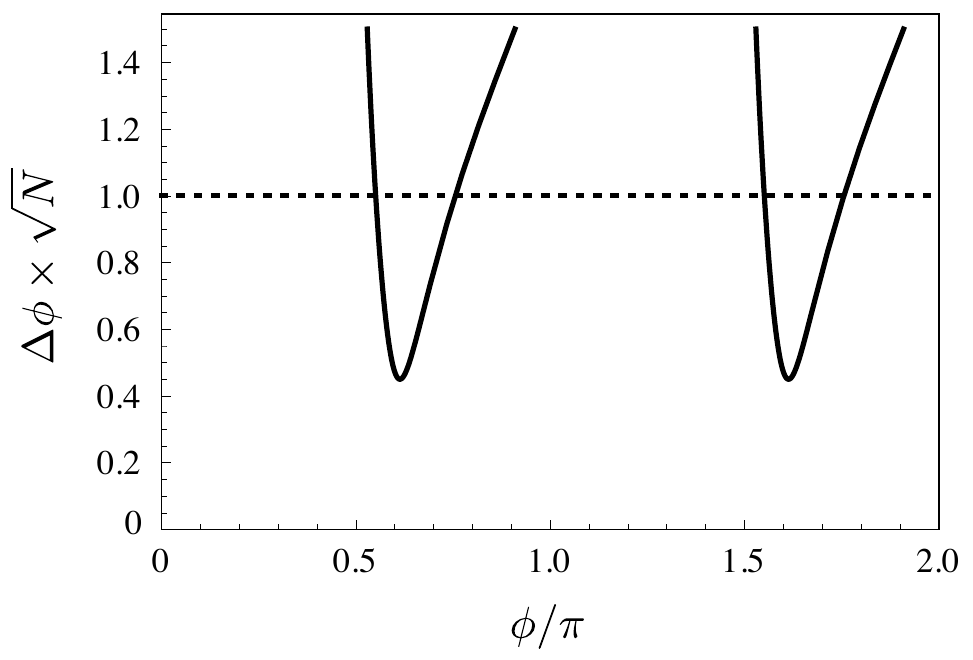}{Theoretical calculation of normalized phase sensitivity as a function of the phase of the second interferometer. The solid line shows the case where an on-axis twisting scheme has been used to prepare quantum squeezing, the dotted line shows the standard quantum limit. The preparation time of the interferometer was chosen as $20\units{ms}$. All parameters are consistent with the experimental set up presented in this paper.}


\section{Conclusions and Outlook}
\label{sect:conclusion}

We envision a sensor in which a squeezed state is created in trap and then the atom cloud is released into free fall to minimise dephasing. Consider, for example, an atom based gravimeter such as the laser cooled system operating in Paris \cite{bodart10}. A sample of atoms could be produced as in the current experiment and released. The falling cloud can be split, reflected and recombined with Raman beams. We calculate that for the large numbers of atoms used in our experiments, the bare, unmodified scattering lengths of the $a_{11}$, $a_{22}$ and $a_{12}$ states will lead to a significant and usable amount of squeezing in $10-20\units{ms}$. The squeezed state could then be used as the source for the atom gravimeter. In such a system we calculate a single shot accuracy of $10^{-9}g$ for a free fall time of $100\units{ms}$, with a further factor to be gained from sub-shot noise statistics. Such a system, based on BEC, is currently under construction in our laboratory. 

In conclusion, we have presented results from a trapped, large atom number interferometer operating on the $\hf{F=1}{m_F=0} \rightarrow \hf{F=2}{m_F=0}$ transition in \Rb{87}. We have demonstrated that simple absorption imaging with an inexpensive CCD camera can have a sensitivity of $10\units{dB}$ beyond the standard quantum limit for $10^6$ atoms. We have discussed the principal sources of technical and fundamental noise in our system which must be overcome to achieve projection-noise-limited performance, and have suggested techniques for minimising these.


\section{Acknowledgements}

The authors acknowledge financial support from the Australian Research Council Centre of Excellence program. We thank ANU Summer Scholars Shaun Johnstone and Sam Hile for their work with the spin flips. We are also very grateful to Colin Dedman for his work on the power supply servo. Specific product citations are for the purpose of clarification only, and are not an endorsement by the authors, ANU, UQ, Monash University, JQI or NIST.


\bibliography{asn}
\bibliographystyle{unsrt}

 \end{document}